# Imaging the Energy Gap Modulations of the Cuprate Pair Density Wave State

Zengyi Du[1,*], Hui Li[1,2,*], Sang Hyun Joo[1,3], Elizabeth P. Donoway[1,4], Jinho Lee[3], J. C. Séamus Davis[5,6], Genda Gu[1], Peter D. Johnson[1] and Kazuhiro Fujita[1,**]

[1] CMPMS Department, Brookhaven National Laboratory, Upton, NY 11973, USA
[2] Department of Physics and Astronomy, Stony Brook University, Stony Brook, NY 11790, USA
[3] Department of Physics and Astronomy, Seoul National University, Seoul 08826, Republic of Korea
[4] Department of Physics, University of California, Berkeley, CA 94720, USA
[5] Department of Physics, University College Cork, Cork T12R5C, Ireland
[6] Clarendon Laboratory, University of Oxford, Oxford, OX1 3PU, UK
\* These authors contributed equally to this work
\*\* Corresponding author. Email: kfujita@bnl.gov (K. F.)

**The defining characteristic [1,2] of Cooper pairs with finite center-of-mass momentum is a spatially modulating superconducting energy gap $\Delta(r)$. Recently, this concept has been generalized to the pair density wave (PDW) state predicted to exist in cuprates[3,4]. Although the signature of a cuprate PDW has been detected in Cooper-pair tunneling[5], the distinctive signature in single-electron tunneling of a periodic $\Delta(r)$ modulation has never been observed. Here, using a new approach, we discover strong $\Delta(r)$ modulations in $Bi_2Sr_2CaCu_2O_{8+\delta}$ that have eight-unit-cell periodicity or wavevectors $Q \approx 2\pi/a_0\,(1/8, 0);\, 2\pi/a_0\,(0, 1/8)$. Simultaneous imaging of the local-density-of-states $N(r, E)$ reveals electronic modulations with wavevectors $Q$ and $2Q$, as anticipated when the PDW coexists with superconductivity. Finally, by visualizing the topological defects in these $N(r, E)$ density waves at $2Q$, we discover them to be concentrated in areas where the PDW spatial phase changes by $\pi$, as predicted by the theory of half-vortices in a PDW state[6,7]. Overall, this is a compelling demonstration, from multiple single-electron signatures, of a PDW state coexisting with superconductivity in the canonical cuprate $Bi_2Sr_2CaCu_2O_{8+\delta}$.**

**1** The exact nature of the cuprate pseudogap state[8] has been the focus of extensive research as a route to understanding high temperature superconductivity. Attention has recently focused on a PDW state[3,4] as a leading candidate to be the fundamental order parameter that characterizes the pseudogap. This was originally motivated by transport studies[9], which led to the hypothesis of "stripe superconductivity", in which the superconducting order parameter is spatially modulated and thus a PDW[10,11]. Equally, the highly unusual band structure reconstruction at $T^*$ as observed by angle resolved photoemission spectroscopy[12] (ARPES) can be explained relatively simply by the formation of a PDW state[13,14]. Indeed, a wide variety of microscopic theories based on strong, local electron-electron interactions now envisage a cuprate PDW state[15,16,17-22], while experimental evidence for its existence is rapidly emerging from multiple techniques[4,5,23,24].

**Characteristic Signatures in Single-electron Tunneling of the PDW State**

**2** Here we focus on the challenge of detecting the cuprate PDW state using single-electron tunneling. First, we consider a PDW, whose spatially dependent energy-gap is $\Delta(\mathbf{r}) = F_P \Delta_\mathbf{Q}^P [e^{i\mathbf{Q}\cdot\mathbf{r}} + e^{-i\mathbf{Q}\cdot\mathbf{r}}]$, where $\Delta_\mathbf{Q}^P$ is the amplitude of gap modulations at wavevector $\mathbf{Q}$, and $F_P$ the form factor with either *s*- or *d*-symmetry. The most obvious and immediate prediction is that the single-electron tunneling should detect a gap in the density-of-states spectrum $N(E)$, which modulates at $\mathbf{Q}$. It is striking, therefore, that no such modulating $\Delta(\mathbf{r})$ has ever been observed in the cuprates. Second, if such a PDW coexists with *d*-wave superconductivity (SC), whose homogeneous gap is $\Delta^S(\mathbf{r}) = F_{SC}\Delta^S$, where $F_{SC}$ exhibits *d*-symmetry, then a Ginzburg-Landau (GL) theory predicts the form of $N(\mathbf{r}, E)$ modulations generated by the interactions between the PDW $\Delta(\mathbf{r})$ and the superconducting $\Delta^S(\mathbf{r})$. These modulations are identifiable from products of these two order parameters that transform as density-like quantities. Thus, considering the product of the PDW and SC order parameters, $\Delta_\mathbf{Q}^P \Delta^{S*}$ predicts $N(\mathbf{r}) \propto \cos(\mathbf{Q}\cdot\mathbf{r})$ modulations at the PDW wavevector $\mathbf{Q}$, while the product of a PDW with itself $\Delta_Q^P \Delta_{-Q}^{P*}$ predicts $N(r) \propto \cos(2Q.r)$ at twice the PDW wavevector. Therefore, a second unique signature of a PDW with wavevector $Q$ in the superconducting cuprates would be the coexistence of two sets of $N(r, E)$ modulations at $Q$ and $2Q$. Finally, a topological defect with $2\pi$ phase winding[25] in the induced density wave $N(r) \propto \cos(2Q \cdot r)$ is predicted to generate a local phase winding of $\pi$ in the PDW order, at a half vortex[6] (Fig. 1a). This is the topological signature of a PDW coexisting with homogeneous superconductivity. Experimental detection of these phenomena in single-electron tunneling would constitute compelling evidence for the PDW state.

**3** To explore these predictions, we use a spectroscopic imaging scanning tunneling microscopy[26] (SI-STM) with a $Bi_2Sr_2CaCu_2O_{8+\delta}$ nanoflake tip[5] to visualize the single electron tunneling. Utilization of the superconducting tip enhances the energy-resolution due to the convolution of spectra that are sharply peaked at superconducting gap edge, in the density-of-states $N_T(E)$ of the tip and $N(r, E)$ of the sample. Thus, energy-sensitivity to modulations in $\Delta(r)$ should be enhanced with this superconductor-insulator-superconductor (SIS) tunneling technique. To enable it, a bulk single crystal sample of $Bi_2Sr_2CaCu_2O_{8+\delta}$ at the hole density $p \sim 0.17 \pm 0.01$ and superconducting transition temperature $T_c$ = 91 K is cleaved at room temperature under ultra-high vacuum condition ($3 \times 10^{-10}$ Torr), and then inserted into the cryogenic STM head. The superconducting tip is created by picking up a nanometer scale $Bi_2Sr_2CaCu_2O_{8+\delta}$ flake from the sample[5] to form the SIS junction. The SI-STM measurements throughout this paper are then all performed using such SIS junctions at $T$ = 9 K. A typical SIS topography is shown in Fig. 1b for a 40 nm × 40 nm field-of-view (FOV). Individual Bi atoms in the BiO plane with sub-atomic resolution are resolved as shown in the inset. The $CuO_2$ plane exists $\sim$ 6 Å below the BiO plane.

**Direct visualization of the periodic energy gap modulations**

**4** Differential SIS conductance spectra $g(r, E) \equiv dI/dV(r, E = eV)$ are then measured as a function of position in this FOV for the energy range from −150 meV to +150 meV. A typical such spatially averaged $g(r, E)$ spectrum is shown in red in Fig. 1c, together with a normal-insulator-superconductor metal (NIS) spectrum earlier performed on the same sample but in a different FOV. The SIS $g(r, E)$ spectrum being a convolution of the tip $N_T(\epsilon)$ and sample $N(\epsilon + E)$ demonstrates enhanced energy resolution as expected (red Fig. 1c). Here, since the spatially averaged NIS $g(r,E)$ spectrum is peaked at ±37 meV while the equivalent SIS spectrum peaks at ±66 meV, we estimate the average energy gap of the tip $\Delta_T$ to be 29 meV.

**5**   Next, by measuring half the magnitude of the energy which separates the SIS spectrum peaks at every location, and then subtracting $\Delta_T$, we determine the local gap energy map $\Delta(r)$ in the sample. A typical example is shown in Fig. 2a. Figure 2b shows the magnitude of the power-spectral-density Fourier transform $\Delta(q)$ of $\Delta(r)$ from Fig. 2a. Equivalent results have been achieved using SIS tunneling with three different $Bi_2Sr_2CaCu_2O_{8+\delta}$ nanoflake tips, on three different samples, and with two different STMs (Methods Section C.1). In Fig. 2b, $q_{SM}$ corresponds to a wavevector of the crystal-structure supermodulation. This supermodulation does indeed produce a type of PDW detectable by its energy gap modulations; but this PDW is trivial, occurring due to modulation of the crystal unit-cell dimensions (Methods Section C.3). Second there is a very broad peak in $\Delta(q)$ surrounding $q = 0$ due to the well-known random energy-gap disorder of $Bi_2Sr_2CaCu_2O_{8+\delta}$, and this is equivalent to the broad range of gap values in the histogram inset to Fig. 2a. Finally, there are four distinct local maxima in $\Delta(q)$ at the points indicated by black solid dots surrounding $q \approx (0, \pm 0.125); (\pm 0.125, 0) 2\pi/a_0$.

**6**   These features indicate that there is a strong, if disordered, modulation in $\Delta(r)$, running parallel to the Cu-O-Cu bonds of the $CuO_2$ plane. This modulation exists on top of a non-periodic energy gap $\approx 37$ meV. It exhibits well-defined peaks at $Q_x \approx 2\pi/a_0$ (1/

8,0) and $Q_y \approx 2\pi/a_0$ (0,1/8) meaning that $\Delta(r)$ is modulating with $\approx 8a_0$ periodicity along both axes. Such a variation in $\Delta(r)$ can be seen directly in a series of SIS $g(r,E)$ spectra, extracted along the line in Fig. 2a and shown in Fig. 2c. Here we see a local demonstration of how that the energy of maximum $N(r)$ (i.e. of the coherence peak) is modulating at $\approx 8a_0$ periodicity in a particle-hole symmetric fashion with an amplitude of approximately 6 meV. More fundamentally, line profiles from $\Delta(q)$ in Fig. 2b are plotted in Fig. 2d for both directions. The two well-defined peaks in Fig. 2d characterize a PDW with wavevectors $Q_x = (0.129 \pm 0.003, 0); Q_y = 2\pi/a_0 (0, 0.118 \pm 0.003)$. This is the first observation of a coherent modulation in the superconducting energy gap $\Delta(r)$ in any material, and is precisely what is expected for a PDW state. Moreover, it reveals directly that the cuprate PDW occurs at wavevectors $Q \approx 2\pi/a_0 (1/8,0); 2\pi/a_0 (0,1/8)$.

**Relationship to PDW visualization using Scanned Josephson Tunneling Microscopy**

7     Using the same $Bi_2Sr_2CaCu_2O_{8+\delta}$ nanoflake tip technology, on samples at the same doping as herein but operating at millikelvin temperatures, the magnitude of the Josephson current $|I_J(r)|$ is found to modulate with wavelength $\approx 4a_0$. Thus, modulations of $|I_J(r)|$ and of $\Delta(r)$ are both detectable when using nanoflake tips that are extracted from the same crystal that is being studied, and are likely in the same coexisting SC and PDW state. Because the nanoflake tip is extended, an approximation to planar tunneling must be considered. Here $I_J$ from an extended tip to the crystal is composed of two contributions: $I_J^S$ due to pair tunneling from $\langle c_k^\dagger c_{-k}^\dagger \rangle$ to $\langle c_k^\dagger c_{-k}^\dagger \rangle$ states, and $I_J^P$ due to pair tunneling from $\langle c_k^\dagger c_{-k+Q}^\dagger \rangle$ to $\langle c_k^\dagger c_{-k+Q}^\dagger \rangle$ states, which are independent of each other when pair-momentum is conserved (Methods Section C2). In scanned Josephson tunneling microscopy, the circuitry measures the magnitude of Josephson critical current magnitude: $|I_J| = |I_J^S| + |I_J^P|$, for which $|I_J^S|$ is a roughly constant spatially but $|I_J^P| \propto |Sin(Q_P\delta)|$ where $\delta$ is the spatial displacement between the PDW in the extended tip and the PDW in the sample (Methods

Section C2). Under these circumstances, if the PDW has periodicity $8a_0$ its gap modulates with periodicity $8a_0$, but the magnitude of the total Josephson current $|I_J|$ will have periodicity $4a_0$. This is the specific phenomenology detectable using the $Bi_2Sr_2CaC_2O_{8+\delta}$ nanoflake tips for SIS spectroscopy and to measure the magnitude of Josephson critical currents[5], respectively. Further, enhanced sensitivity to the basic energy modulations when using SIS spectroscopy is consistent with a 'lock-in' effect from a PDW state in the nanoflake tip (Methods Section C4).

**Detection of two unidirectional PDWs within distinct nanoscale domains**

*8*      Next, in order to explore the unidirectionality of $\Delta(\boldsymbol{r})$, we employ a two-dimensional lock-in technique to determine the amplitude and phase of the modulations[27]. Thus

$$A_{\boldsymbol{Q}}(\boldsymbol{r}) = \int d\boldsymbol{R} A(\boldsymbol{R}) e^{i\boldsymbol{Q}\cdot\boldsymbol{R}} e^{-\frac{(r-R)^2}{2\sigma^2}} \qquad (1)$$

$$|A_{\boldsymbol{Q}}(\boldsymbol{r})| = \sqrt{ReA_{\boldsymbol{Q}}(\boldsymbol{r})^2 + ImA_{\boldsymbol{Q}}(\boldsymbol{r})^2} \qquad (2)$$

$$\Phi_{\boldsymbol{Q}}^A(\boldsymbol{r}) = tan^{-1}\frac{ImA_{\boldsymbol{Q}}(\boldsymbol{r})}{ReA_{\boldsymbol{Q}}(\boldsymbol{r})}, \qquad (3)$$

where $A(\boldsymbol{r})$ represents any arbitrary real space image, $\boldsymbol{Q}$ the wavevector of interest, and $\sigma$ the averaging length-scale in $r$-space (or equivalently the cut-off length in $q$-space). The key ingredients of such an analysis are the amplitude $|A_{\boldsymbol{Q}}(\boldsymbol{r})|$ and the spatial phase $\Phi_{\boldsymbol{Q}}^A(\boldsymbol{r})$ of moduations at $\boldsymbol{Q}$. Using this technique on our $\Delta(\boldsymbol{r})$ data, Figure 3a,b show the amplitudes of the PDW for the $x$ and $y$ directions, $|\Delta_{\boldsymbol{Q}_x}(\boldsymbol{r})|, |\Delta_{\boldsymbol{Q}_y}(\boldsymbol{r})|$, respectively. The local "magnitude" of PDW unidirectionality is then defined as

$$F(\boldsymbol{r}) = \frac{|\Delta_{\boldsymbol{Q}_x}(\boldsymbol{r})| - |\Delta_{\boldsymbol{Q}_y}(\boldsymbol{r})|}{|\Delta_{\boldsymbol{Q}_x}(\boldsymbol{r})| + |\Delta_{\boldsymbol{Q}_y}(\boldsymbol{r})|}. \qquad (4)$$

When $F(\mathbf{r})>0$, represented in orange, the PDW along the $x$-direction is dominant, and similarly when $F(\mathbf{r})<0$, represented in blue, the PDW order along the $y$-direction is dominant. As shown in Fig. 3c, $F(\boldsymbol{r})$ is spatially heterogeneous forming a domain structure indicating that the cuprate PDW $\Delta(r)$ is microscopically unidirectional, with one direction predominant in any particular domain. In addition, it appears that the domain size in orange is significantly bigger than that of blue within the 40nm x 40nm FOV, which may indicate a vestigial nematic[28] PDW state, although one cannot be certain without independent knowledge of the shape anisotropy of the nanoflake tip. Overall, these data indicate that the cuprate PDW state is locally strongly unidirectional, and possibly in a vestigial nematic state due to quenched disorder[28].

**How a coexisting PDW and superconductor induce the CDW modulations**

**9**     Although the SI-STM technique cannot be used to image a charge density $\rho(r)$ or any of its modulations, a mapping of $g(r,E)$ and its ratio $Z(r,E)=g(r,+E)/g(r,-E)$ enables one to study how the related $N(r,E)$ modulates. It has been found that the form-factor symmetry for the induced CDW in cuprates exhibits primarily $d$-symmetry[27]. In that case, the CDW modulation does not appear primarily at $Q$ and $2Q$ in the Fourier transform of $g(r,E)$ or $Z(r,E)$. Instead, to detect the $d$-symmetry form factor CDW signal at $Q$ and $2Q$, one must first use the $d$-symmetry sublattice-phase-resolved Fourier analysis (Methods Section D.1). For this reason, we apply a phase-resolved visualization of the $d$-symmetry modulations to our measured $Z(r,E)$[27], extracting the value of $Z(r,E)$ at the oxygen sites within each CuO$_2$ unit cell: $O_x^Z(r) \equiv Z(r)\delta(r - r_{O_x})$ at O$_x$ and similarly for $O_y^Z(r)$ at O$_y$. We then subtract these values throughout the image to yield

$$D^Z(r) = O_x^Z(r) - O_y^Z(r). \qquad (5)$$

In Fig. 4a and b we show the amplitudes of $|\Delta_{Q_x}(r)|$ and $|D_{2Q_x}^Z(r, E = 54 \text{ meV})|$; systematics of the $q$-space cutoff length used are discussed in Methods Section D2. If we then consider the magnitude of the Fourier transform of $D^Z(r,E)$ for $E$ = 54 meV where SIS tunneling has the maximum energy-sensitivity (Fig. 1c), a key fact emerges. In the Fourier transform $|D^Z(q, 54 \text{ meV})|$ we find two strong peaks at $Q$ and $2Q$ (Fig. 4c), which are the clearest in these data when presented along the line $(0,0) - 2\pi/a_0 (0.4,0)$ in Fig. 4d (from which a Lorentzian background has been subtracted). This complex density wave structure is the expected signature in $N(r, E)$ modulations[29,30,31] of the PDW with wavevector $Q$ coexisting with the homogeneous superconductivity.

**10**    Finally, utilizing the two-dimensional lock-in technique to generate phase-resolved images, the spatial phase $\Phi^Z_{2Q_x}(r)$ of $D^Z_{2Q_x}(r, 54\text{ meV})$ is extracted using Eqn. 3, and is shown in Fig. 4e. The topological defects with $2\pi$ phase winding in the $N(r, 54\text{ meV})$ density wave are marked by the black and white dots, for which the winding direction is clockwise and counter clockwise, respectively. The presence of these $2\pi$ topological defects in the $N(r, 54\text{ meV})$ density wave at $2Q$, is due microscopically to a dislocation as schematically shown in Fig. 1a (black line). In order to visualize the interplay of the $2Q$ density wave and the PDW, the spatial phase $\Phi^\Delta_{Q_x}$ of the PDW order is extracted in the same way, but now at $2\pi/a_0\ (\pm 1/8, 0)$. In Fig. 4f, the locations of the $\Phi^Z_{2Q_x}(r)$ topological defects from Fig. 4e are also plotted on top of the PDW spatial phase $\Phi^\Delta_{Q_x}(r)$. Intriguingly, the $\Phi^Z_{2Q_x}(r)$ topological defects are always found in the vicinity of the yellow strings in $\Phi^\Delta_{Q_x}(r)$, where the PDW phase is $\pi$ (see Extended Fig. 8 for an evolution of the PDW spatial phase). The inset in Fig. 4f shows that the probability distribution of the PDW phase $\Phi^\Delta_{Q_x}$, at which all the topological defects in $\Phi^Z_{2Q_x}(r)$ are found, is clearly centered around $\pi$ (see Extended Fig. 7 for an independent analysis yielding the same conclusion.). Thus, the local phase in the PDW surrounding the topological defects in $\Phi^Z_{2Q_x}(r)$ always changes by $\sim\pi$ (see Extended Fig. 8) precisely as expected when a topological defect in the induced density wave at $2Q$ interacts with the PDW order[6].

**Multiple single-electron signatures of a PDW coexisting with superconductivity**

**11**    To summarize, use of $Bi_2Sr_2CaCu_2O_{8+\delta}$ nanoflake scanned-tips allows detection of spatially modulating energy-gap $\Delta(r)$ with eight unit-cell periodicity, or with axial wavevectors $\boldsymbol{Q} \approx 2\pi/a_0(1/8, 0); 2\pi/a_0(0, 1/8)$, in superconducting Bi$_2$Sr$_2$CaCu$_2$O$_{8+\delta}$ (Fig. 2). The spatial analysis of the $\Delta(\boldsymbol{r})$ modulations shows that they are rather unidirectional within nanoscale domains (Fig. 3). Simultaneous density-of-states imaging reveals two pairs of co-existing $N(\boldsymbol{r}, E)$ modulations, at wavevectors $\boldsymbol{Q} \approx 2\pi/a_0(1/8, 0); 2\pi/a_0(0, 1/8)$ and $2\boldsymbol{Q} \approx 2\pi/a_0(1/4, 0); 2\pi/a_0(0, 1/4)$ (Fig. 4c and d). Finally, the topological defects in the $N(\boldsymbol{r}, E)$ density wave at $2\boldsymbol{Q}$ are concentrated along the lines where the PDW spatial phase changes by π (Fig. 4f). All of these phenomena are canonical signatures[4,6,28,30,31] of a PDW coexisting with homogeneous superconductivity. Thus, $\Delta(\boldsymbol{r})$ modulation imaging provides direct spectroscopic evidence of the existence of a PDW, at zero magnetic field in cuprates.

**Acknowledgements**: We thank S. D. Edkins, E. Fradkin, M. H. Hamidian, S. A. Kivelson, P. A. Lee and J. M. Tranquada, for erudite discussions and excellent advice. Z.D., H.L. and K.F. G.G. and P.D.J. acknowledge support from the U.S. Department of Energy, Office of Basic Energy Sciences, under contract number DEAC02-98CH10886. S.H.J. acknowledges support from the Institute for Basic Science in Korea (Grant No. IBS-R009-G2), the Institute of Applied Physics of Seoul National University and National Research Foundation of Korea (NRF) grant funded by the Korea government (MSIP) (No. 2017R1A2B3009576). E.P.D. was supported by the Brookhaven National Laboratory Supplemental Undergraduate Research Program (SURP). J.C.S.D. acknowledges support from Science Foundation of Ireland under Award SFI 17/RP/5445 and from the European Research Council (ERC) under Award DLV-788932.

**Author Contributions:** K.F. designed and lead the project. Z.D., H.L., S.H.J., E.P.D., and K.F. carried out experiments at the MIRAGE STM of the OASIS complex at Brookhaven Nat. Lab.; G.G. synthesized and characterized the samples; Z.D., H.L. and K.F. developed and carried out analysis. K.F. wrote the paper with key contributions from J.C.S.D, Z.D., H.L., J.L., and P.D.J. The manuscript reflects the contributions and ideas of all authors.

**Author Information** Reprints and permissions information is available at www.nature.com/reprints. The authors declare no competing financial interests. Readers are welcome to comment on the online version of the paper. Correspondence and requests for materials should be addressed to kfujita@bnl.gov

**FIGURES**

**Figure 1. Schematic of unidirectional $8a_0$ PDW order intertwined with density wave. a,** Unidirectional PDW order parameter is modulated along the horizontal axis at eight unit-cell periodicity. Sign of $\Delta(r)$ is colored in red for positive and in blue for negative. The periodicity in the unidirectional density-of-electronic-states $N(r)$ detectable by NIS tunneling, which is intertwined with unidirectional PDW, is depicted by a black broken line. $N(r)$ wavelength is half that of unidirectional PDW. When a dislocation occurs in unidirectional $N(r)$, where the $2\pi$ phase winding is realized in its phase, a possible fluctuation in unidirectional PDW order amplitude and/or half-vortices are predicted. The relative phase of the spatial variations in the PDW order and the induced density wave modulations in $N(r)$ are also plotted at the top and bottom of a. **b,** Typical SIS topography of $Bi_2Sr_2CaCu_2O_{8+\delta}$ within 40 nm x 40 nm field-of-view (FOV) at 5 GOhm junction resistance ($I$ = 30 pA at $V_{bias}$ = 150 meV). Cu-O-Cu bond directions are in parallel to $x$ and $y$ axes. Individual Bi atoms are clearly observed as shown in the inset with intra-unit-cell resolution. **c,** Spatially averaged SIS $g(r)$ spectrum (red) together with that taken with NIS junction (black). Spatially averaged SIS $g(r,E)$ spectrum shows particle-hole symmetric peaks with energies at $E$ = ± 66 meV. Since spatially averaged NIS $g(r)$ is peaked at ± 37 meV, we estimate the average gap value on the $Bi_2Sr_2CaCu_2O_{8+\delta}$ nanoflake tip to be about 29 meV.

**Figure 2. Superconducting gap energy modulations. a,** Measured $\Delta(r)$ within 40 nm × 40 nm FOV. The energy-gap data presented here and throughout the manuscript are all the measured values of the magnitude of $\Delta(r)$, which is half the energy separation between the coherence peaks minus the gap value of the $Bi_2Sr_2CaCu_2O_{8+\delta}$ nanoflake tip. The inset shows a distribution of measured $\Delta(r)$ in the same FOV, ranging from 27 meV to 56 meV. The eight unit-cell modulation is clearly resolved in $\Delta(r)$, primarily running along the $x$ direction of Cu-O-Cu bond exhibiting the unidirectional signature of the PDW. **b,** The magnitude Fourier transform of **a**. Well-defined Fourier peaks at $Q \cong \frac{2\pi}{a_0}(\pm 1/8, 0); \frac{2\pi}{a_0}(0, \pm 1/8)$, corresponding to the eight unit-cell modulations for both x and y directions, are observed.  **c,** The series of SIS g(r,E) spectra intensity along the line in **a**. The eight unit-cell modulation of the energies of the coherence peaks is clearly resolved. The modulation amplitude is about 6 meV. **d,** The line cut of |Δ(**q**)| along both x and y directions from b. As seen in **b**, $Q \cong \frac{2\pi}{a_0}(\pm 1/8, 0); \frac{2\pi}{a_0}(0, \pm 1/8)$ peaks are present for both directions, but a peak height is about twice larger for x than that for y direction indicating that the PDW is rather unidirectional.

**Figure 3. The PDW order parameter amplitude and phase. a,** The spatial variation of the PDW amplitude $|\Delta_{Q_x}(r)|$ along the x-direction obtained by the two-dimensional lock-in technique.  **b,** The spatial variation of the PDW amplitude $|\Delta_{Q_y}(r)|$ along the y direction obtained by the two-dimensional lock-in technique. The cut-off length is denoted by the broken circle. **c,** The local directionality map defined by Eq. (4). An orange domain is the region where the PDW amplitude along the x axis is predominant, while a blue domain is the region where the PDW amplitude along the y axis is predominant.

**Figure 4.  The interplay of N(r) and PDW, and the possible half-vortices. a,** The spatial variation of the N(r) amplitude $D^Z_{2Q_x}(r)$ obtained by Eq. (1) and (2). Inset shows an azimuthal angular averaged cross correlation coefficient as a function of distance **b,** The spatial variation of the $\Delta_{Q_x}(r)$ amplitude obtained by Eq. (1) and (2). **c,** The magnitude of the phase-resolved Fourier transform, $|D^Z(q)|$ exhibiting both $Q = 2\pi/a_0(\pm 1/8, 0)$ and $2Q = 2\pi/a_0(\pm 1/4, 0)$ peaks encircled by red broken lines, respectively. Coordinates are in units of $2\pi/a_0$. **d** The line cut of $|D^Z(q)|$, in which Lorentzian background is subtracted, in **c** from $2\pi/a_0(0,0)$ to $2\pi/a_0(\pm 1/4, 0)$, exhibiting well-defined peaks at **Q** and 2**Q**. Data points are fitted by Lorentzians and the obtained

peak positions are $2\pi/a_0(0.113 \pm 0.002)$ and $2\pi/a_0(0.241 \pm 0.003)$ for **Q** and **2Q**, respectively, with the peak widths to be $2\pi/a_0(0.016 \pm 0.004)$ and $2\pi/a_0(0.068 \pm 0.006)$ for **Q** and **2Q**, respectively. **e,** The spatial phase of the 2**Q** N(**r**) order $\Phi^Z_{2Q_x}(r)$ obtained by eq. (3). $2\pi$ topological defects are marked by solid dots. White (black) dots indicate $2\pi$ phase winding along clockwise (counter clockwise). **f,** The spatial phase of the PDW $\Phi^\Delta_{Q_x}(r)$. The $2\pi$ topological defects in $\Phi^Z_{2Q_x}(r)$ from **e** are plotted on top of $\Phi^\Delta_{Q_x}(r)$. An inset shows the distribution of $\Phi^\Delta_{Q_x}(r)$ values at all the locations where the $2\pi$ topological defects in $\Phi^Z_{2Q_x}(r)$ are found.

# METHODS

## A. Sample Preparation and Measurement

*A.1. Sample preparation*

High-quality $Bi_2Sr_2CaCu_2O_{8+\delta}$ single crystals were grown by the traveling-solvent floating zone method. Here we measured a sample with hole doping level of $p \approx 0.17$. The sample, covered by the Kapton tape, was loaded into the load-lock chamber with pressure better than $3\times10^{-10}$ Torr and quickly inserted into the STM head at T≈9K, after cleavage by removing the Kapton tape.

## B. Tip preparation and characterization

*B.1. Tip isotropy*

The tip isotropy is checked by comparing the height of the Bragg peaks for x- and y-directions in the Fourier transform $T(\boldsymbol{q})$ of topographic images using the nanoflake tip $T(\boldsymbol{r})$. A 40nm x 40nm FOV $T(\boldsymbol{r})$ and its real-part Fourier transform $\text{Re}T(\boldsymbol{q})$ are shown in ED Fig. 1a and 1b, respectively. ED Fig. 1c shows line profiles of $\text{Re}T(\boldsymbol{q})$ along the lines in ED Fig. 1b across the Bragg peaks at (1,0) and (0,1). Bragg peak heights at (1,0) and (0,1) are found to be comparable within 7%.

## C. Motivation and model for $\Delta(r)$ modulation detection

*C.1. Motivation of searches for a PDW signature in $\Delta(r)$*

Here, we discuss preliminary $\Delta(r)$ data, as shown in ED Fig. 4a and Fig. 4c, that motivated and provides reinforcement for the data presented in this study, which is completely independent of them. These data in ED Fig. 4 were acquired with two different Bi2212 nanoflake tips, on two different samples, using two different SISTM instruments. Although experimental conditions were not optimized for detection of $\Delta(r)$ modulations in a PDW, the peaks in $\Delta(q)$ at $Q = (0, \pm0.125); (\pm0.125, 0)2\pi/a_0$ are weakly visible as marked by dashed white circles in ED Fig. 4b and ED Fig. 4d. Such data, along with those reported in the main text from an experiment designed and optimized for the purpose, provide the type of experimental evidence available on the existence of $8a_0$ modulations in $\Delta(r)$.

*C.2. Independent Pair Tunneling Process*

Here we discuss how the $4a_0$ modulation observed in the magnitude of the Josephson critical current and the $8a_0$ modulation observed in $\Delta(r)$ in the present study may be linked. We consider a simple model for pair tunneling from a nanoflake Bi2212 tip that is in the coexisting SC and PDW state, to the parallel surface of a bulk Bi2212 crystal in the same state (ED Fig. 3). Because the tip is extended and parallel to the surface, the effects of planar tunneling must be considered. In perfect planar tunneling, the $\langle c_k^\dagger c_{-k}^\dagger \rangle$ Cooper pairs of the homogenous SC cannot tunnel into the $\langle c_k^\dagger c_{-k+Q}^\dagger \rangle$ Cooper pairs of the PDW, because that violates conservation of momentum. In that limit, the Josephson current $I_J$ from a Bi2212 extended tip to the Bi2212 crystal is composed of two independent contributions: $I_J^S$ due to pair tunneling from $\langle c_k^\dagger c_{-k}^\dagger \rangle$ to $\langle c_k^\dagger c_{-k}^\dagger \rangle$ states, and $I_J^P$ due to pair tunneling from $\langle c_k^\dagger c_{-k+Q}^\dagger \rangle$ to $\langle c_k^\dagger c_{-k+Q}^\dagger \rangle$ states.

Consider two PDW, one in the nanoflake tip $\Psi_T$ and one in the sample $\Psi_S$, with order parameters

$$\Psi_T = \Delta_1 e^{i\theta_1}\left(e^{i(Q(x+\delta))}\right) \; ; \; \Psi_S = \Delta_2 e^{i\theta_2}\left(e^{iQx}\right)$$

The Josephson coupling will be of the form

$K\left(\Psi_T \Psi_S^* + \Psi_T^* \Psi_S\right) \propto cos(\theta_1 - \theta_2)cos(Q\delta)$ where $K$ is a constant and $\delta$ is the variable spatial displacement of the tip PDW relative to the sample PDW. In this case, the inter PDW Josephson current takes the form

$$I_J^P \propto \sin(\theta_1 - \theta_2)\sin(Q\delta)$$

It is the magnitude $|I_J^P|$ that is measured as a function of transverse displacement $\delta$ between nanoflake tip and sample where $Q = 2\pi/\lambda$, and this obviously modulates as $|I_J^P| \propto |sin(Q\delta)|$ or with a periodicity of $\lambda/2$.

Our previous studies using scanned Josephson tunneling (*Nature 571*, 541(2016)) actually measured the magnitude of the Josephson current $|I_J|$. Thus, if $I_J^S$ and $I_J^P$ are independent, then $|I_J| = |I_J^S| + |I_J^P|$. Assuming that $|I_J^S|$ is roughly constant spatially, then $|I_J^P| \propto |sin(Q\delta)|$ where $\delta$ is the transverse displacement between the PDW in the extended tip and the PDW in the sample. Therefore, in this model for our experiment, if the PDW has true periodicity $8a_0$ then its gap modulation $\Delta(r)$ will necessarily have periodicity $8a_0$ but, critically, the modulations in magnitude of the total Josephson current $|I_J|$ will have periodicity $4a_0$.

Note that if there are two strictly independent unidirectional PDWs with wavevectors $\boldsymbol{Q}_x$ and $\boldsymbol{Q}_y$, and Cooper pair momentum of each is conserved, then the $\boldsymbol{Q}_x$ PDW cannot tunnel to the $\boldsymbol{Q}_y$ PDW and vice versa. This would pose a challenge to the above analysis. However, if the PDW state in the tip is somewhat biaxial (e.g. *Proc. Natl'. Acad. Sci.* **115**, 5389-5391 (2018)), then this analysis would retain validity.

*C.3. Effect of structural supermodulation on measured $\Delta(r)$.*

One might ask if there is an effect of the crystal supermodulation with $\boldsymbol{Q_{SM}} \| (1,1)2\pi/a_0$ which produces an energy gap modulation at its wavevector, on our measured $\Delta(r)$. As we reported in Ref. 5, we observed the modulations both in $\Delta(r)$ and Josephson critical current at $\boldsymbol{Q_{SM}}$. However, this is a trivial effect and its wavevector is at 45 degrees off the Cu-O-Cu direction. Most importantly, a spatial convolution between the tip and sample of their modulating $\Delta(r)$ at $\boldsymbol{Q}_{SM}$ cannot produce any additional modulations at different wavevectors. Thus, the effect of structural supermodulation does not produce any other gap modulation signals, especially at $\boldsymbol{Q} = (0, \pm 0.125); (\pm 0.125, 0)2\pi/a_0$.

*C.4. Effect of $\Delta(r)$ on nanoflake*

Here, we discuss how $\Delta(r)$ modulation detection is enhanced in BSCCO nanoflake SIS tunneling. Here, the measured $\Delta(r)$ can be regarded as a consequence of a spatial convolution between the sample and nanoflake PDW order parameters. The nanoflake is most likely in the same PDW state since it is picked up from the same sample. Thus, the order parameter on the nanoflake is approximated in a form of $\Delta_{nanoflake} exp(i\boldsymbol{Q_P} \cdot \boldsymbol{r}) exp(-\frac{r^2}{2\sigma^2})$, where the exponential term is approximated to represent a nanoflake structure factor with size of nanoflake (~3nm, see ED. Fig.2). This acts as a low-pass filter in the convolution between gap modulations at the same wavevector $\boldsymbol{Q_P}$ in the tip and in the sample. Such a convolution effect naturally works as a "lock-in", mitigating the signals unrelated to the gap modulation wavevector $\boldsymbol{Q_P}$. This process makes the signal of $\Delta(r)$ modulation detectable.

***D. Data Analysis***

*D.1. Sub-lattice phase-resolved analysis*

To reveal any possible local-density-of-states $N(\boldsymbol{r},E)$ modulations, we analyze differential conductance $g(\boldsymbol{r},E)$ to yield $Z(\boldsymbol{r}, E)=g(\boldsymbol{r}, +E) / g(\boldsymbol{r}, -E)$[26]. Intensities at Oxygen sites $\boldsymbol{r}_{O_x}$ and $\boldsymbol{r}_{O_y}$ are extracted separately from $Z(\boldsymbol{r}, E=54\text{meV})$ and used to form two new maps, $O_x^Z(\boldsymbol{r}, E = 54\text{mV})$ and $O_y^Z(\boldsymbol{r}, E = 54\text{meV})$, respectively. We then calculate each sublattice-phase-resolved $Z(\boldsymbol{r}, E)$ image and separate it into three: the first, $Cu(\boldsymbol{r})$, contains only the measured values of $Z(\boldsymbol{r})$ at Cu sites while the other two, $O_x(\boldsymbol{r})$ and $O_y(\boldsymbol{r})$, contain only the measurements at the *x/y*-axis oxygen sites.

Phase-resolved Fourier transforms of the $O_x(\boldsymbol{r})$ and $O_y(\boldsymbol{r})$ sublattice images $\tilde{O}_x(\boldsymbol{q}) = Re\tilde{O}_x(\boldsymbol{q}) + iIm\tilde{O}_x(\boldsymbol{q})$; $\tilde{O}_y(\boldsymbol{q}) = Re\tilde{O}_y(\boldsymbol{q}) + iIm\tilde{O}_y(\boldsymbol{q})$, are used to determine the form factor symmetry for modulations at any $\boldsymbol{q}$

$$\tilde{D}^Z(\boldsymbol{q}) = (\tilde{O}_x(\boldsymbol{q}) - \tilde{O}_y(\boldsymbol{q}))/2$$

$$\tilde{S}'^{Z}(\boldsymbol{q}) = (\tilde{O}_x(\boldsymbol{q}) + \tilde{O}_y(\boldsymbol{q}))/2$$

$$\tilde{S}^{Z}(\boldsymbol{q}) = \widetilde{Cu}(\boldsymbol{q})$$

Specifically, for a DW occurring at $\boldsymbol{Q}$, one can then evaluate the magnitude of its $d$-symmetry form factor $\tilde{D}(\boldsymbol{Q})$ and its $s'$- and $s$-symmetry form factors $\tilde{S}'(\boldsymbol{Q})$ and $\tilde{S}(\boldsymbol{Q})$, respectively. In terms of the segregated sub-lattices, a $d$-form factor DW is one for which the DW on the $O_x$ sites is in anti-phase with that on the $O_y$ sites. Studies of electronic structure in underdoped $Bi_2Sr_2CaCu_2O_{8+x}$ and $Ca_{2-x}Na_xCuO_2Cl_2$ consistently exhibit a relative phase of $\pi$ and therefore a $d$-symmetry form factor.

Hence the peaks at $\pm\boldsymbol{Q_x}$ and $\pm\boldsymbol{Q_y}$ present in both $\tilde{O}_x(\boldsymbol{q})$ and $\tilde{O}_y(\boldsymbol{q})$ must cancel exactly in $\tilde{O}_x(\boldsymbol{q}) + \tilde{O}_y(\boldsymbol{q})$. Therefore, if a density wave at $\boldsymbol{Q}$ and $\boldsymbol{2Q}$ has predominantly $d$-symmetry form factor, there is no detectable signal in $g(\boldsymbol{r},E)$ or $Z(\boldsymbol{r}, E)$ at $\boldsymbol{Q}$ and $\boldsymbol{2Q}$, and why the d-symmetry Fourier transform $D^g(\boldsymbol{q},E)$ or $D^Z(\boldsymbol{q},E)$ are used in these studies. Specifically, by calculating $D^Z(\boldsymbol{q}) = \text{FFT}(D^Z(\boldsymbol{r}))$ one correctly extracts the $d$-symmetry density wave modulations that are occurring at $\boldsymbol{Q}$ and $\boldsymbol{2Q}$.

*D.2. Cut-off dependence*

Here, we show how the images shown in Fig. 4 evolve as a function of cut-off length used in the two-dimensional lock-in technique. In E.D. Fig. 6 both $D^Z_{2Q_x}(\boldsymbol{r}, 54\text{meV})$ and $\Delta_{Q_x}(\boldsymbol{r})$ are shown at different real-space cut-off lengths, 8 Å, 16Å, 24Å, 32Å, and 40Å. On the left column, we can see a big change between 8 Å and 16Å in the spatial structure of $\left|D^Z_{2Q_x}(\boldsymbol{r})\right|$ as oscillatory components are vanished, while $\left|D^Z_{2Q_x}(\boldsymbol{r})\right|$ at for 16Å, 24Å, 32Å, and 40Å are virtually identical. For $\left|\Delta_{Q_x}(\boldsymbol{r})\right|$ on the right column in E.D. Fig. 6, the oscillatory components are gone between 16 Å and 24Å as. Thus, the cut-off length used in Fig. 3, 16 Å and 40Å, don't introduce erroneous oscillations by picking up irrelevant contributions from other wavevectors and are reasonable choices.

*D.3. Interplay of the eight-unit-cell periodic PDW and the four-unit-cell induced N(r,E) modulation*

In order to support the Fig. 4f inset, in which 2π topological defects in the induced $N(\boldsymbol{r})$ modulation at 2$\boldsymbol{Q}$ tends to be found in the vicinity of the locus of π phase in $\Phi_{Q_x}^{\Delta}(\boldsymbol{r})$ (yellow strings) we performed an independent analysis: the distances of the white and black dots to the nearest position on the yellow strings are calculated and compared to randomly distributed results. ED. Fig. 7a shows the distance distribution of the total 25 topological defects in Fig. 4e. Then we generate randomly distributed 25 "topological defects" inside the same field of view and calculate distances to the same yellow strings, and this process has been repeated 100 times. The average result of the 100 different configurations is shown in E.D. Fig. 7b. It is very clear that the distribution from the measured $N(\boldsymbol{r})$ topological defects at 2$\boldsymbol{Q}$ is in a smaller range with higher magnitude compared to random results. This supports that the topological defects in the measured $N(\boldsymbol{r})$ modulation at 2$\boldsymbol{Q}$ actually show a statistically strong tendency to be found near the locus of π phase in $\Phi_{Q_x}^{\Delta}(\boldsymbol{r})$

*D.4. π phase winding and possible half-vortex in PDW*

In search for half-vortices in PDW, we analyzed PDW phase $\Phi_{Q_x}^{\Delta}(\boldsymbol{r})$ in the vicinity of the 2π topological defects from the induced $N(\boldsymbol{r})$ modulation at 2$\boldsymbol{Q}$. We extracted the values along each contour surrounding the 2π topological defects from the induced $N(\boldsymbol{r})$ modulation at 2$\boldsymbol{Q}$ (ED. Fig. 8a) and plotted an evolution of the PDW phase for each contour in ED. Fig. 8b. Although no singularities that have a π phase winding in $\Phi_{Q_x}^{\Delta}(\boldsymbol{r})$ are found, indeed PDW phases are changing by π along each contour, indicating the presence of possible half-vortices.

**Data Availability** The data that support the findings of this study are available from the corresponding author kfujita@bnl.gov upon reasonable request.

**Extended Data Figure Legends**

**Extended Data Fig. 1**: **Analysis of the tip isotropy**. **a**, Topography $T(r)$ within 40nm x 40nm FOV. **b**, Real part of Fourier transform of $T(r)$. **c**, Line profile Re$T(q)$ along the line in the middle panel, representing nearly equal Bragg peak height (difference is less than 7%).

**Extended Data Fig. 2**: **Estimation of the nanoflake tip geometry**. **a**, autocorrelation of $\Delta(r)$. **b**, line profile measured from center in a is azimuthal-angle averaged. The size of the nanoflake on the tip is estimated from Full width at half maximum and is around 3.3nm.

**Extended Data Fig. 3: Possible process of the Josephson Tunneling**. Schematic representation of planar Josephson tunneling in the presence of two order parameters: homogeneous superconductivity and PDW.

**Extended Data Fig. 4: Preliminary experimental data analysis.** **a**, **c**, Preliminary $\Delta(r)$ measured at 4.2K in nearly optimally doped Bi2212. **b**, **d**, The magnitude of Fourier transform of $\Delta(r)$ in a and c, respectively, representing early observations of 1/8 peaks marked by the red circles.

**Extended Data Fig. 5: Differential conductance map and its ratio. a**, $g(r, E=54$ meV) map. The eight-unit-cell CDW modulation i.e. the PDW induced $N(r)$ modulation at $Q$, can be seen. **b**, $Z(r, E=54$ meV) calculated by $Z(r, E) = g(r, E)/g(r, -E)$.

**Extended Data Fig. 6: Cut-off-length dependence of $|D^Z_{2Q_x}(r)|$ and $|\Delta_{Q_x}(r)|$.** The left column shows $|D^Z_{2Q_x}(r)|$ at different cut-off lengths, similarly for the right column for $|\Delta_{Q_x}(r)|$.

**Extended Data Fig. 7: Distance analysis**. **a,** A count distribution sorted by distances between the topological defects in the induced *N*(*r*) modulation at 2***Q*** from Fig. 4b and the nearest point on the yellow strings in the PDW phase map from Fig. 4c. **b,** Average distribution of 100 configurations, within each configuration 25 points are randomly generated in the same field of view and distances to the same yellow strings are calculated and sorted.

**Extended Data Fig. 8: Spatial evolution of the PDW phase. a,** A phase map $\Phi^\Delta_{Q_x}(r)$ of the PDW order. Three representative contours surrounding the 2π topological defects from $\Phi^Z_{2Q_x}(r)$ across the yellow strings. **b,** An evolution of the phase along each contour in **a**. The upside-down black triangle marks the starting point of winding and the upright black triangle marks the ending point, in correspondence with the winding directions in **a**. π phase windings are clearly seen in the PDW phase surrounding the 2π topological defects from $\Phi^Z_{2Q_x}(r)$.

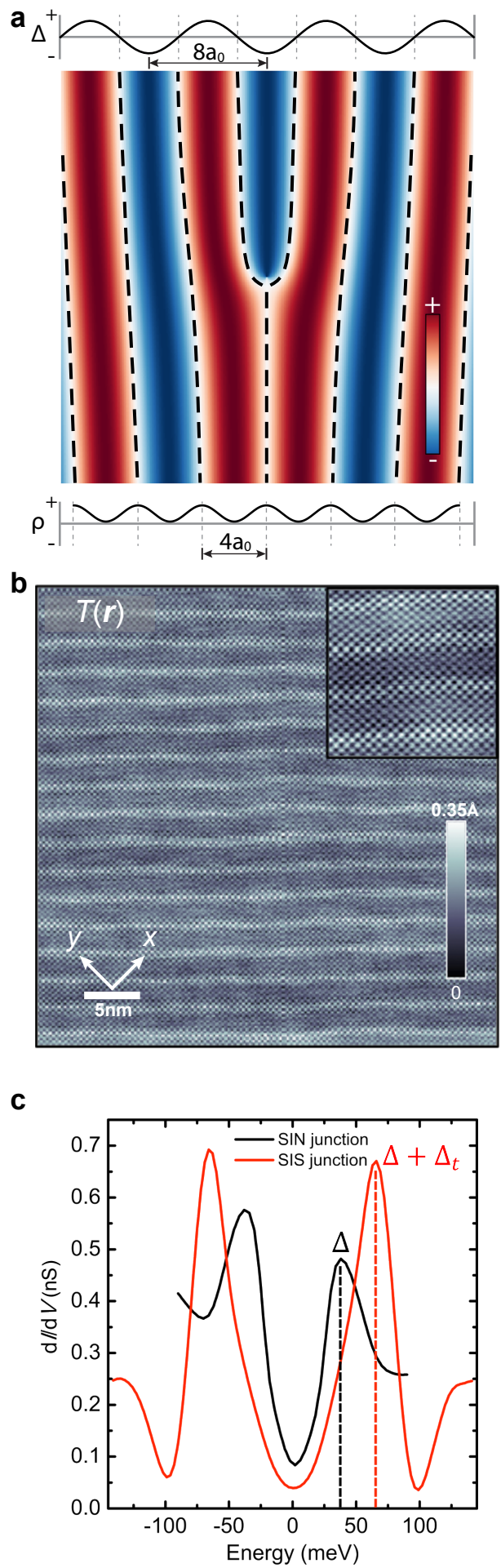

Figure 2

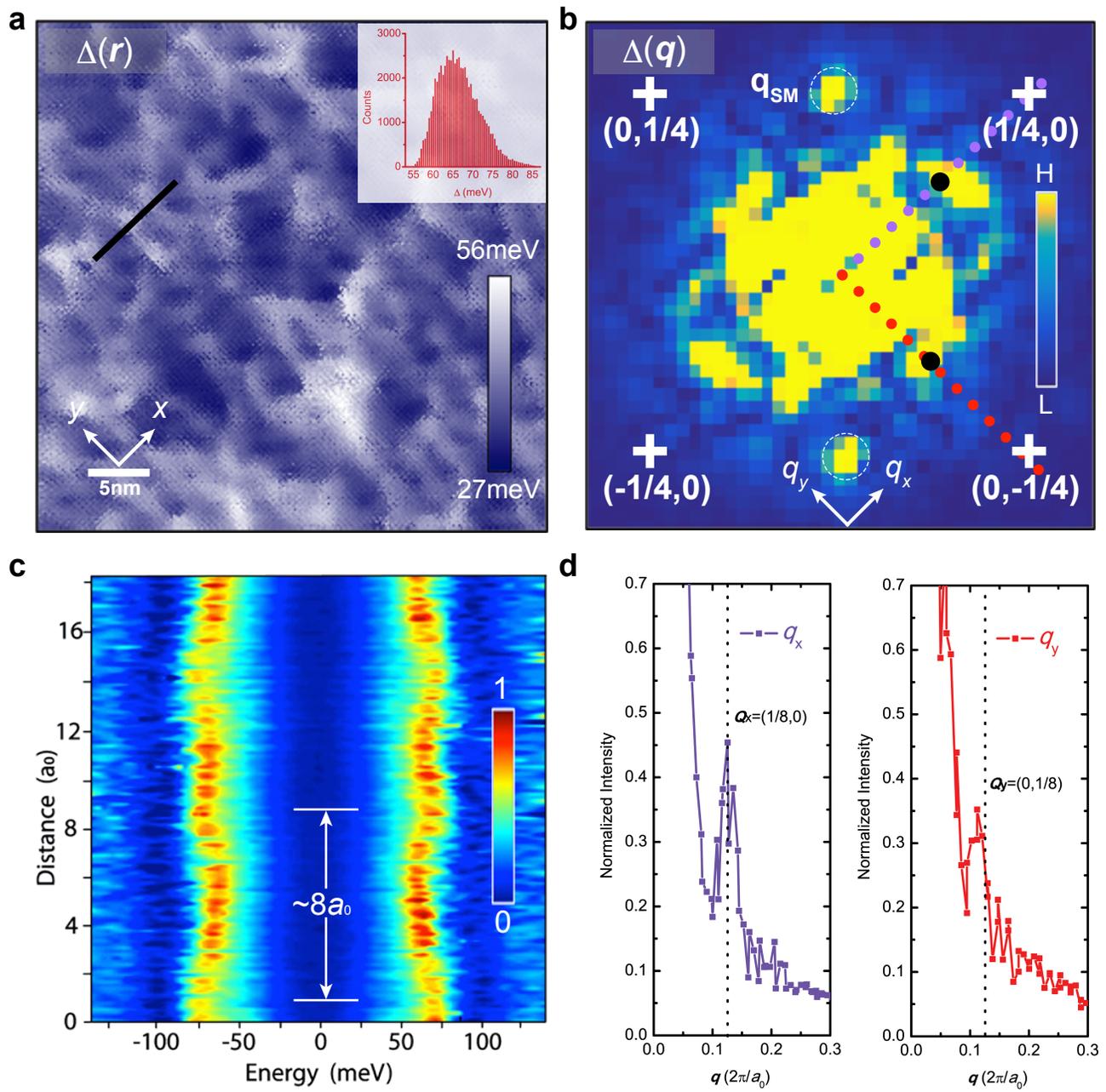

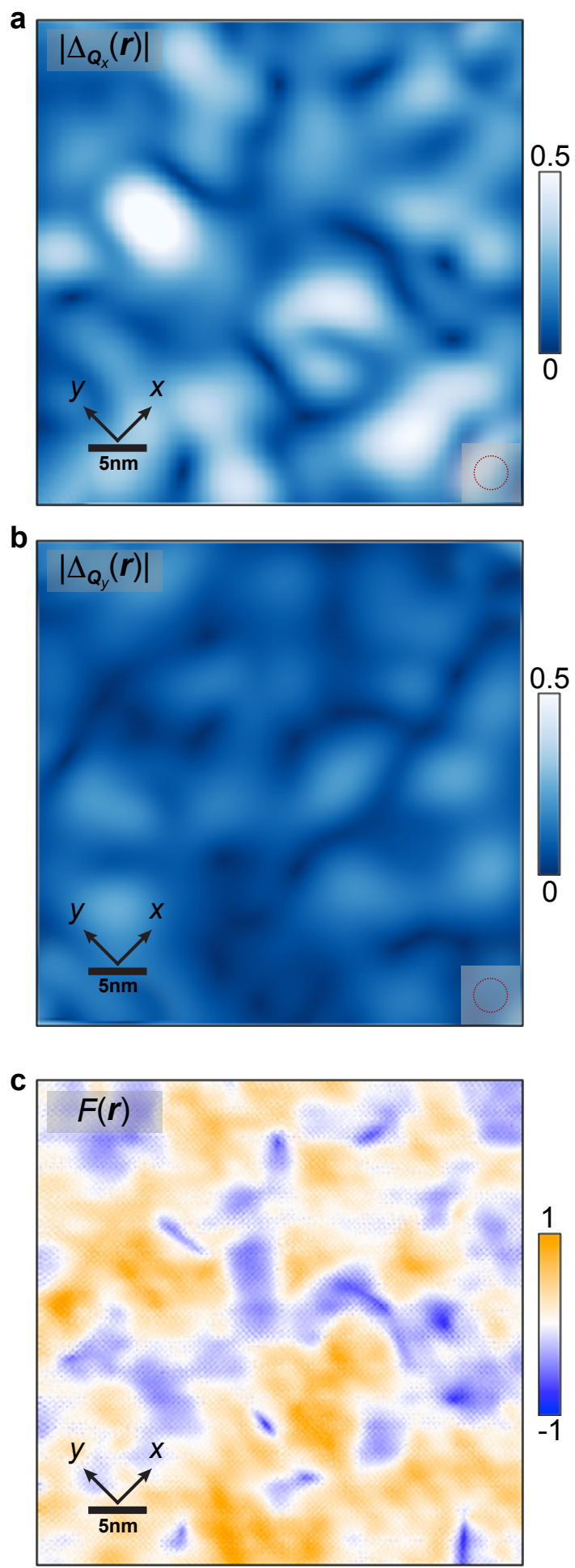

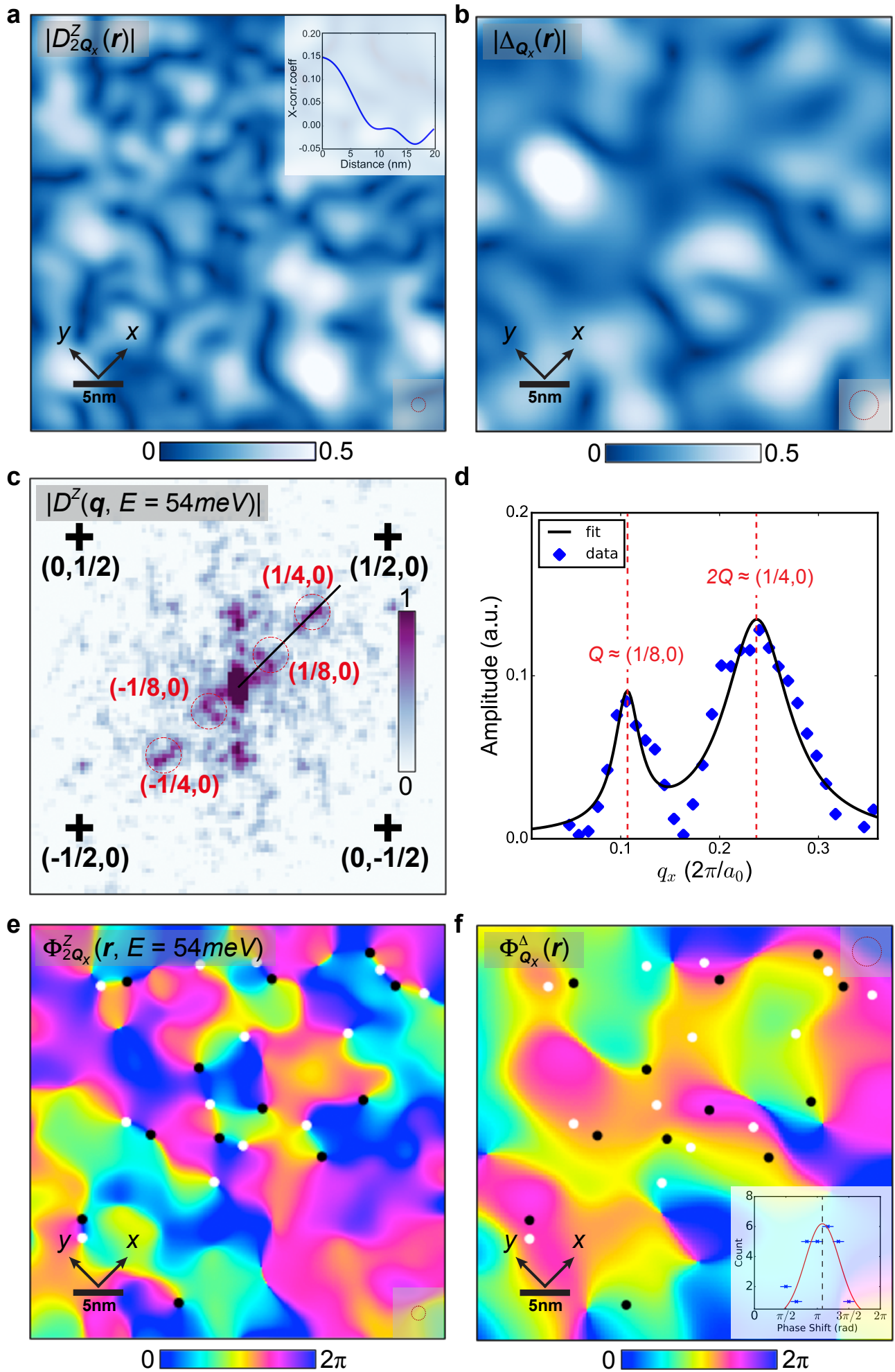

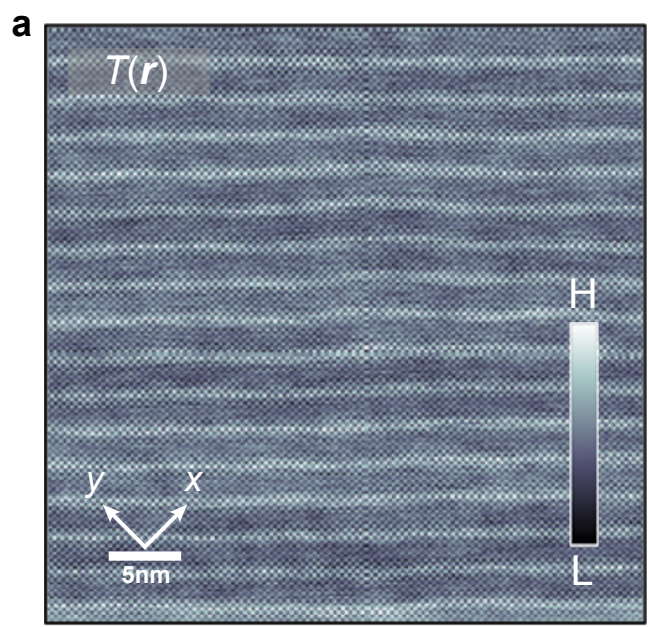
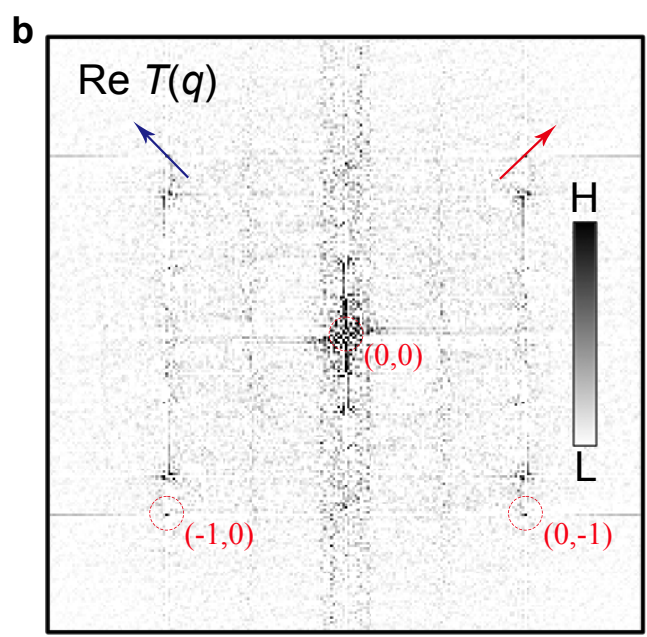
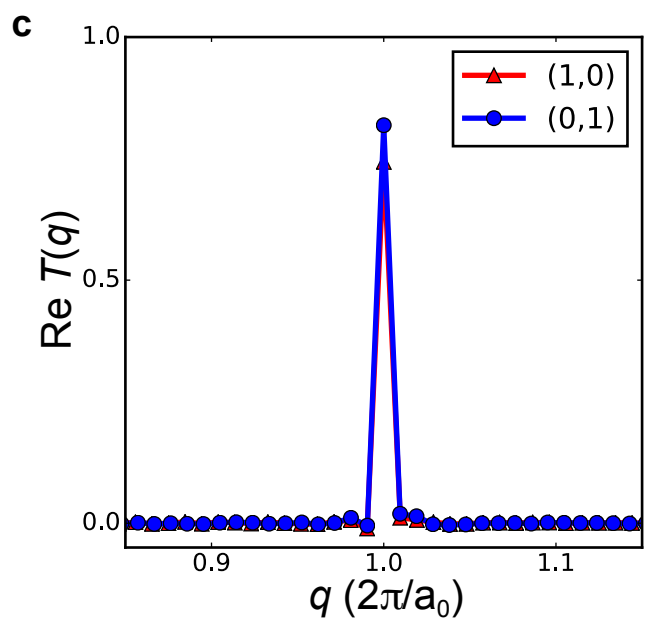

**a**

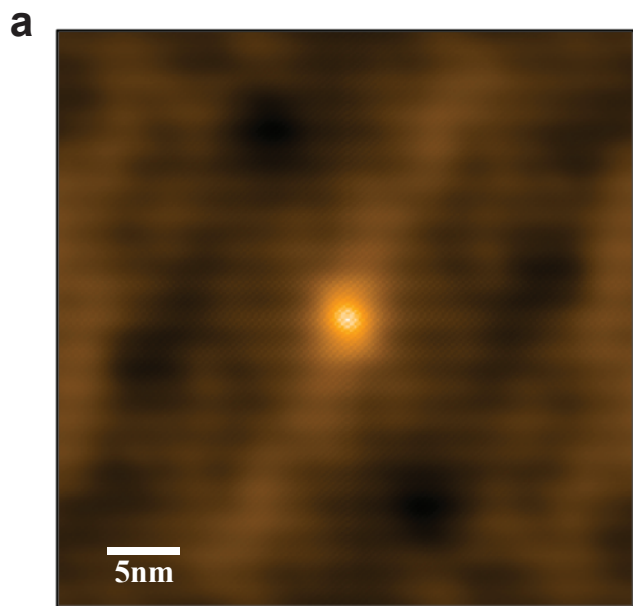

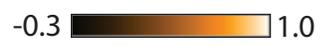
-0.3 ▬▬ ▬▬ 1.0

**b**

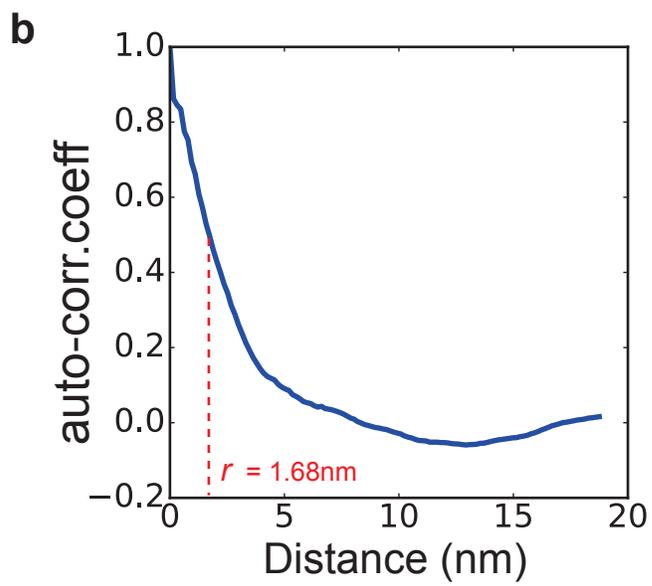

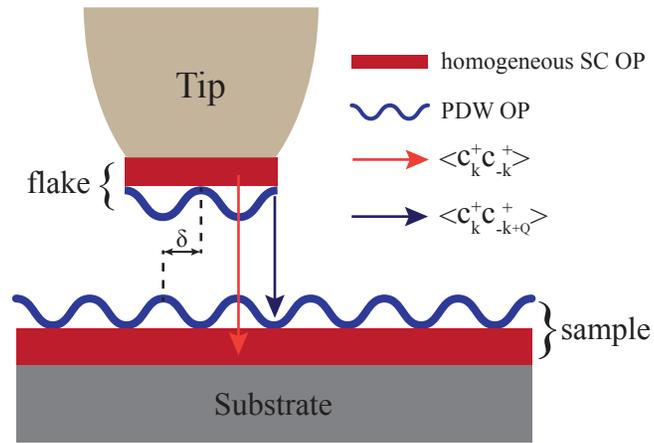

Extended Figure 4

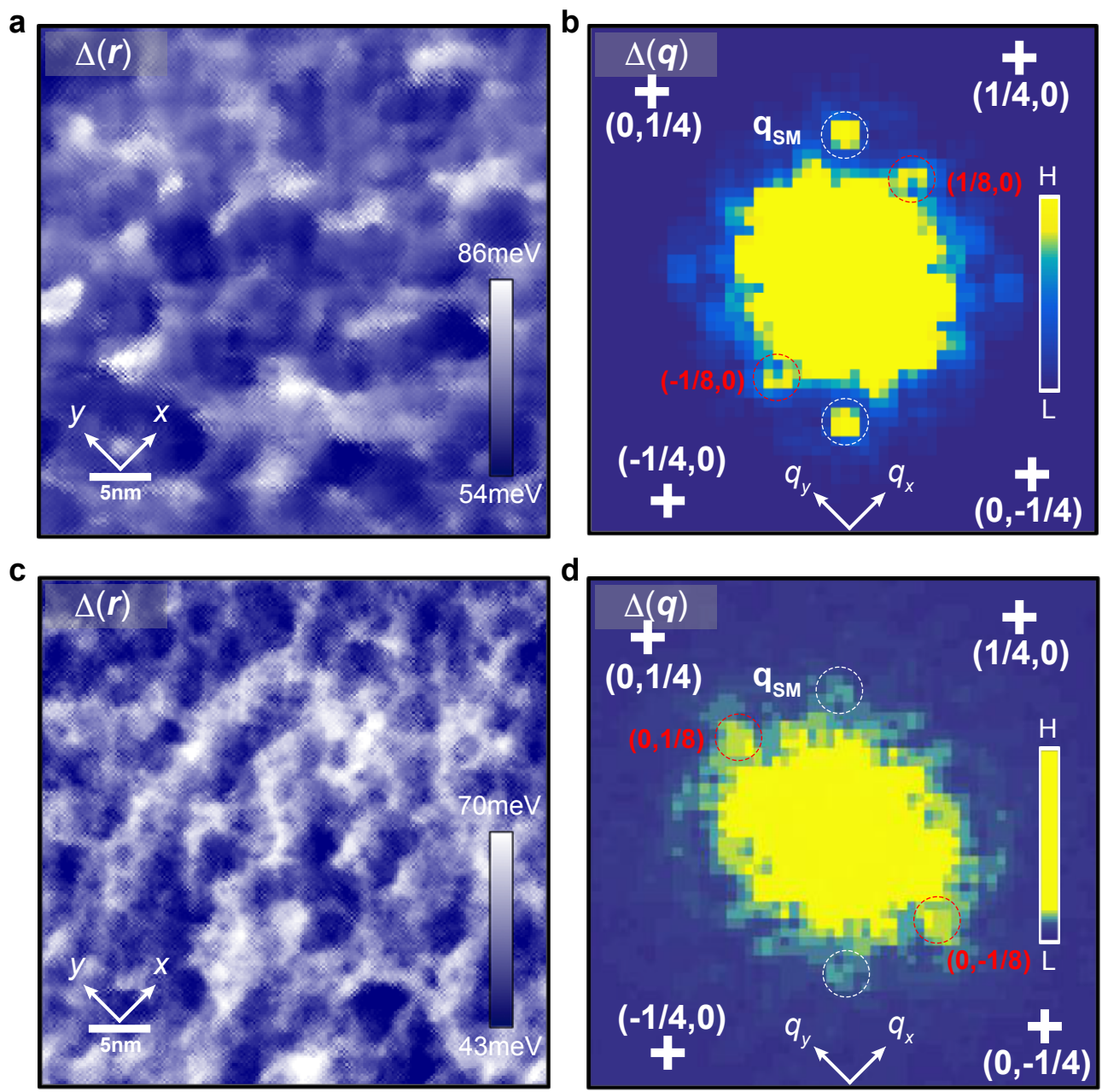

Extended Figure 5

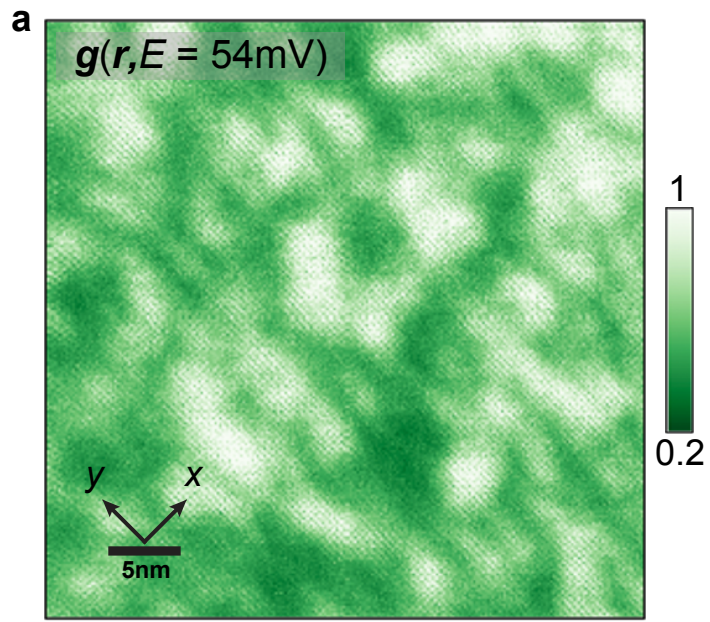

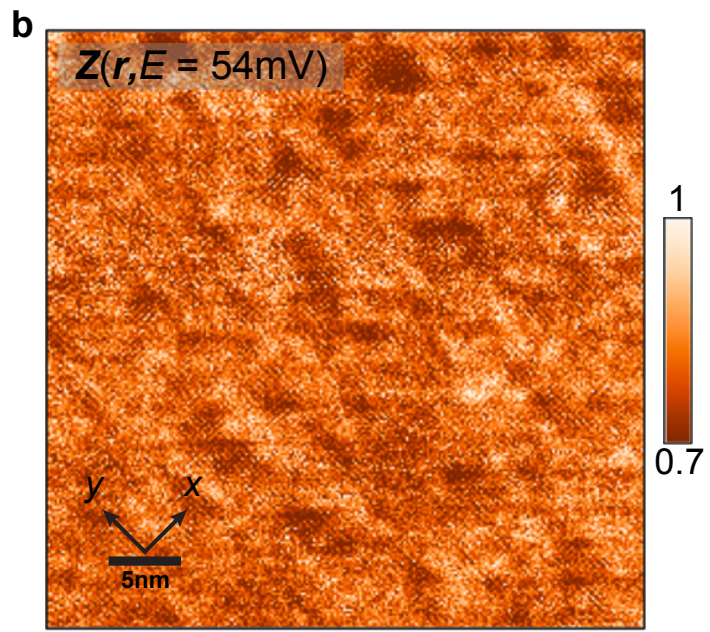

Extended Figure 6

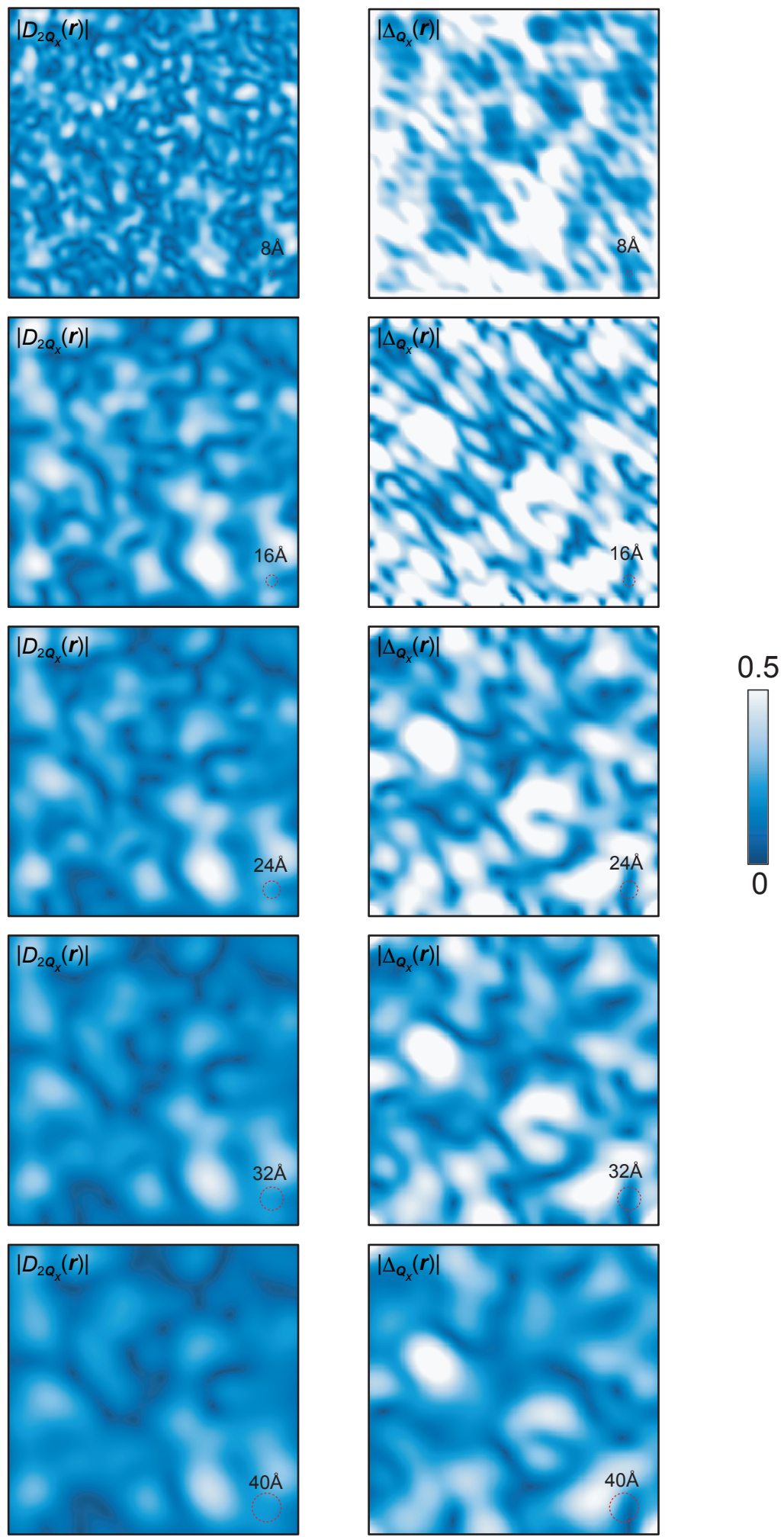

Extended Figure 7

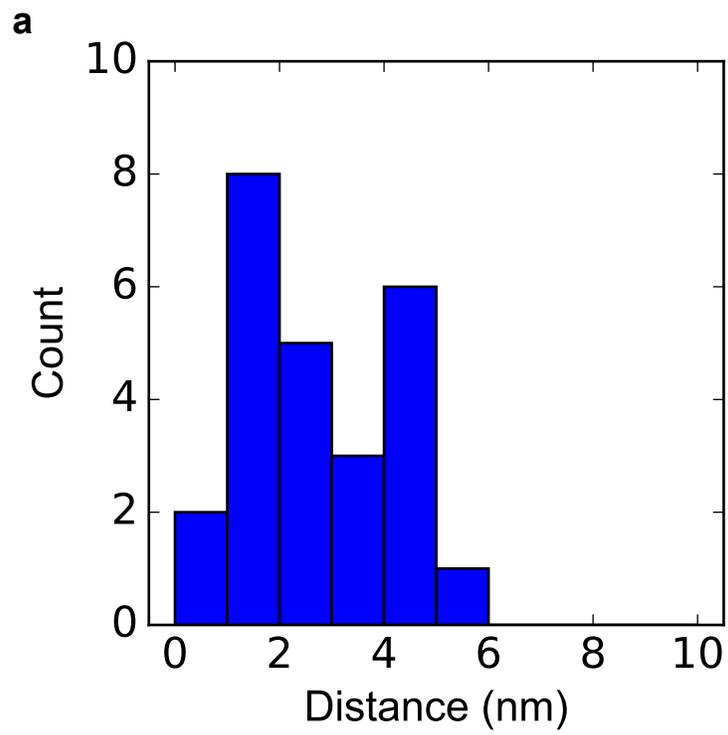

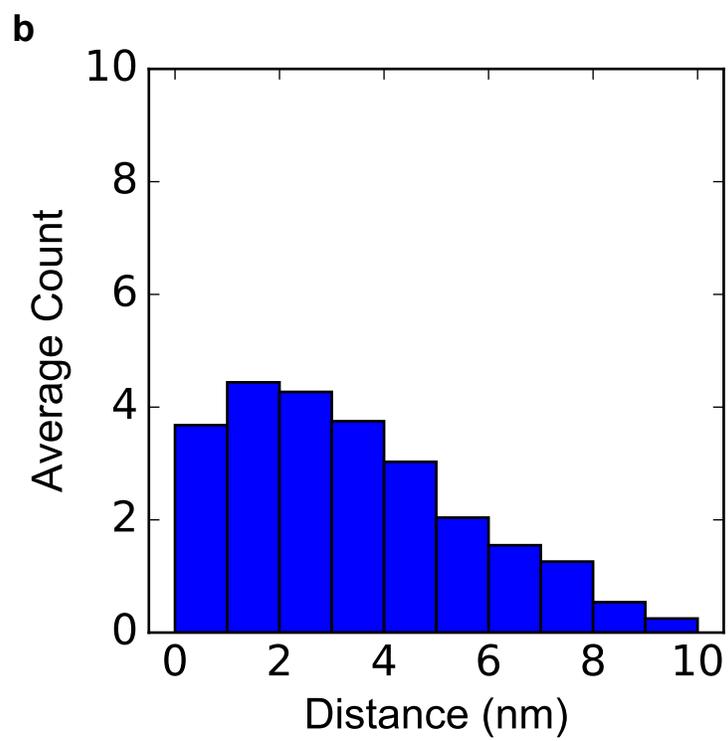

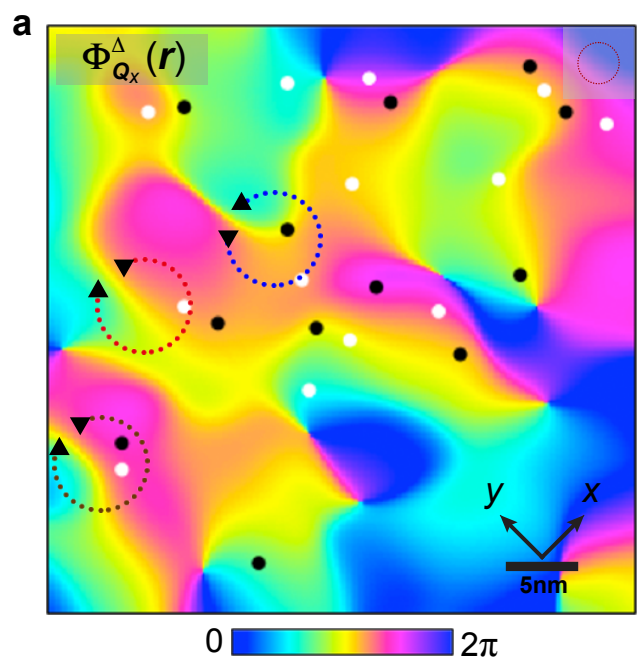

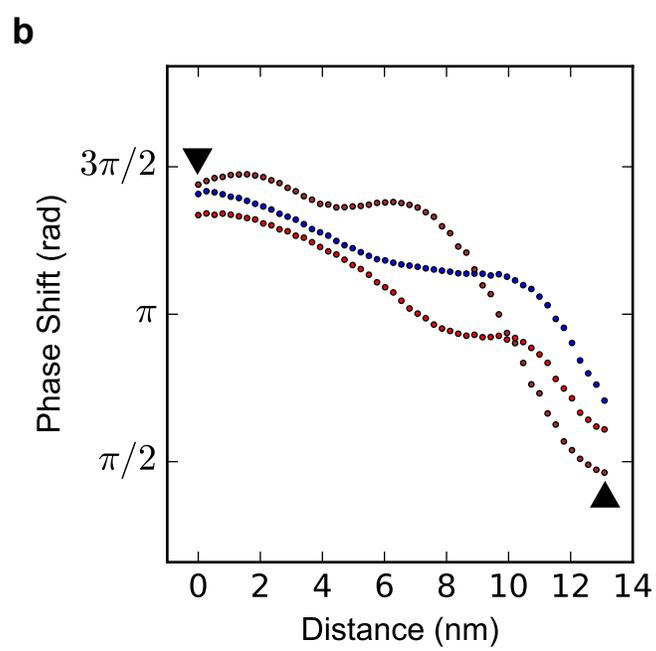